\documentclass[aps,prl,reprint, amsmath, amssymb,superscriptaddress]{revtex4-1}

\pdfoutput=1
\usepackage[utf8]{inputenc}
\usepackage{bm}
\usepackage[english]{babel}
\usepackage[T1]{fontenc}
\usepackage{mathrsfs}
\usepackage[retainorgcmds]{IEEEtrantools}
\usepackage{amsmath}
\usepackage{amssymb}
\usepackage{color}
\usepackage{amsfonts}
\usepackage{times,txfonts}
\usepackage{nicefrac}
\usepackage[colorlinks=true,linkcolor=blue,urlcolor=blue,citecolor=blue,pdfusetitle]{hyperref}

\usepackage{float}
\usepackage{epsfig} 
\usepackage{subfigure}
\graphicspath{{Images/}}

\usepackage{physics}
\usepackage{amssymb}
\usepackage[amssymb]{SIunits}

\usepackage{xcolor}

\newcommand{\trans}{^{\text{T}}}

\begin{document}

\title{Quantum features of entropy production in driven-dissipative transitions}

\author{Bruno O. Goes}
\affiliation{Instituto de Física, Universidade de São Paulo, CEP 05314-970, São Paulo, São Paulo, Brazil}

\author{Carlos E. Fiore}
\affiliation{Instituto de Física, Universidade de São Paulo, CEP 05314-970, São Paulo, São Paulo, Brazil}

\author{Gabriel T. Landi}
\affiliation{Instituto de Física, Universidade de São Paulo, CEP 05314-970, São Paulo, São Paulo, Brazil}

\begin{abstract}
The physics of driven-dissipative transitions 
is currently a topic of great interest, particularly in quantum optical systems. 
These transitions occur in systems kept out of equilibrium and are therefore characterized by a finite entropy production rate. 
However, very little is known about  how the entropy production  behaves around criticality and all of it is restricted to classical systems. 
Using quantum phase-space methods, we put forth a framework that allows for the complete characterization of the entropy production in driven-dissipative transitions.
Our framework is tailored specifically to describe photon loss dissipation, which is effectively a zero temperature process for which the standard theory of entropy production breaks down. 
As an application, we study the open Dicke and Kerr models,  which present continuous and discontinuous transitions, respectively.
We find that the entropy production naturally splits into two contributions. 
One matches the behavior observed in classical systems. The other diverges at the critical point.  

\end{abstract}

\maketitle


{\bf \emph{Introduction - }}
%
The entropy of an open system is not conserved in time, but instead evolves according to 
\begin{equation}
\label{entropy_rate}
    \frac{dS(t)}{dt} = \Pi(t) - \Phi(t),
\end{equation}
where $\Pi \geq 0$ is the irreversible entropy production rate and $\Phi$ is the entropy flow rate from the system to the environment.
Thermal equilibrium is characterized by $dS/dt = \Pi = \Phi = 0$. 
However, if the system is connected to multiple sources, it may instead reach a non-equilibrium steady-state (NESS) where $dS/dt = 0$ but $\Pi = \Phi \geq 0$.
NESSs are therefore characterized by the continuous  production of entropy, which  continuously flows to the environments. 

In certain systems a NESS can also undergo a phase transition.
These so-called dissipative transitions~\cite{Hartmann2006,Diehl2008QuantumAtoms,Verstraete2009} represent the open-system analog of quantum phase transitions.
Similarly to the latter, they are characterized by an order parameter and may be either continuous or discontinuous~\cite{Marro1999,Tomadin2011,Carmichael2015}.
They are also associated with the closing of a gap, although the gap in question is not of a Hamiltonian, but of the Liouvillian generating the open dynamics~\cite{Kessler2012DissipativeSystem,Minganti2018}.
The novel features  emerging from the competition between  dissipation and quantum fluctuations has led to a burst of interest in these systems in the last few years \cite{Kessler2012DissipativeSystem,Lee2013,Torre2013,Minganti2018,Carusotto2013,Sieberer2013,Carmichael2015,Chan2015,Mascarenhas2015a,Sieberer2014,Lee2014a,Weimer2015a,Casteels2016c,Mendoza-Arenas2016,Jin2016b,Bartolo2016a,Casteels2017b,Savona2017a,Rota2017,Biondi2017,Casteels2017,Minganti2018,Raghunandan2018,Gelhausen2018,Foss-Feig2017,Barbosa2018,Lee2018,Hannukainen2017,Gutierrez-Jauregui2018,Vicentini2018,Hwang2017,Vukics2019,Patra2019,Tangpanitanon2019}, including several experimental realizations~\cite{Baumann2010,Landig2015,Fink2016,Fitzpatrick2016,Fink2017,Rodriguez2017,Brunelli2018ExperimentalSystems}.

Given that the fundamental quantity characterizing the NESS is the entropy production rate $\Pi$, it becomes natural to ask how $\Pi$ behaves as one crosses such a transition; i.e., what are its critical exponents? is it analytic? does it diverge? etc.
Surprisingly, very little is known about this and almost all is restricted to classical systems. 

In Refs.~\cite{Tome2012} the authors studied a continuous transition in a 2D classical Ising model subject to two baths acting on even and odd sites. 
They showed that  the entropy production rate was always  finite, but had a  kink at the critical point, with its derivative presenting a logarithmic divergence.
A similar behavior was also observed in  a Brownian system undergoing an order-disorder transition~\cite{Shim2016}, the majority vote model~\cite{Crochik2005}  and a 2D Ising model subject to an oscillating field~\cite{Zhang2016}.
In the system of Ref.~\cite{Zhang2016}, the transition could also become discontinuous  depending on the parameters. 
In this case they found that the entropy production has a discontinuity at the phase coexistence region. 
Similar results have been obtained in Ref.~\cite{Herpich2018} for the dissipated work (a proxy for entropy production) in a synchronization transition. 

All these results therefore indicate that the entropy production is  finite across a dissipative transition, presenting either a kink or a discontinuity.
This general behavior  was recently shown by some of us to be universal for systems described by classical Pauli master equations and  breaking a $Z_2$ symmetry~\cite{Noa2019}. 
An indication that it extends beyond $Z_2$ was given in Ref.~\cite{Herpich2019} which studied a $q$-state Potts model. 

Whether or not this general trend carries over to the quantum domain remains an open question. 
Two results, however, seem to indicate that it does not. 
The first refers to the driven-dissipative Dicke model, studied experimentally in Refs.~\cite{Baumann2010,Landig2015}.
In this system, the part of the entropy production stemming from quantum fluctuations was found to diverge at the critical point~\cite{Brunelli2018ExperimentalSystems}. 
Second, in Ref.~\cite{Dorner2012} the authors studied the irreversible work produced during a unitary quench evolution of the transverse field Ising model.
Although being a different scenario, they also  found a divergence in the limit of zero temperature (which is when the model becomes critical).
Both results therefore indicate that quantum fluctuations may lead to divergences of the entropy production in the quantum regime. 
Whether these divergences are universal and what minimal ingredients they require,  remains a fundamental open question in the field.

\begin{figure}[!t]
    \centering
    \includegraphics[width = 0.3\textwidth]{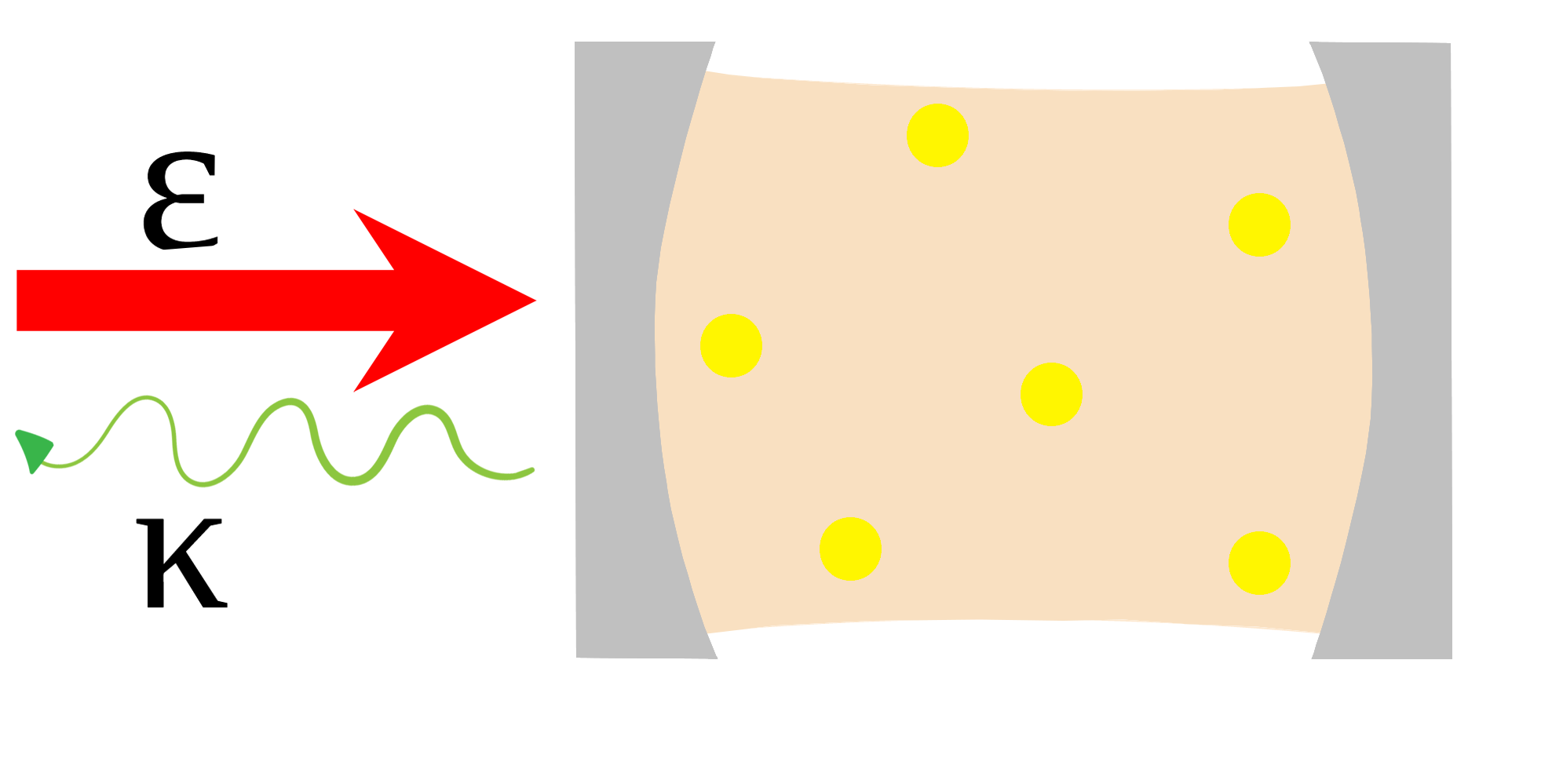}
    \caption{Typical driven-dissipative scenario portraying an optical cavity with a non-linear medium subject to an external pump $\mathcal{E}$ and photon losses occurring at a rate $\kappa$. 
    }
    \label{fig:diagrams}
\end{figure}

The reason why this issue has so far not been properly addressed  is actually technical: most models explored so far fall under the category of a \emph{driven-dissipative process}, where dissipation stems from the loss of photons in an optical cavity~\cite{drummond1980quantum} (see Fig.~\ref{fig:diagrams}).  
The problem  is that photon losses are modeled effectively as a zero temperature bath, for which the standard theory of entropy production yields unphysical results (it is infinite regardless of the state  state or the process)~\cite{Santos2018IrreversibilityEnvironment,Santos2018Spin-phase-space-entropyProduction}. 

This  ``zero-temperature catastrophe''~\cite{Uzdin2018a,Uzdin2019} occurs because the theory relies on the existence of fluctuations which, in classical systems, seize completely as $T \to 0$. 
In quantum systems, however, vacuum fluctuations remain. 
This was the motivation for an alternative formulation  introduced by some of us in Ref.~\cite{Santos2017WignerRate} and recently assessed experimentally in \cite{Brunelli2018ExperimentalSystems}, which uses the Wigner function and its associated Shannon entropy as a starting point to formulate the entropy production problem.
This has the advantage of accounting for the  vacuum fluctuations, thus leading to a framework that remains useful even when $T\to 0$. 

This paper builds on Ref.~\cite{Santos2017WignerRate} to formulate a theory which is suited for describing driven-dissipative transitions.
Since these transitions are seldom Gaussian, we use here instead the Husimi $Q$  function and its associated Wehrl entropy~\cite{Wehrl1978GeneralEntropy,Santos2018Spin-phase-space-entropyProduction}.
Our focus is on defining a consistent thermodynamic limit where criticality emerges. 
This  allow us to separate $\Pi$ into a deterministic term, related to the external laser drive, plus a term related to quantum fluctuations. 
The latter is also additionally split into two terms, one related to the non-trivial unitary dynamics and the other to photon loss dissipation. 
We  apply our results to the Dicke and Kerr models, two paradigmatic examples of dissipative transitions having a continuous and discontinuous transition respectively.
In both cases, we find that unitary part of $\Pi$ behaves exactly like in classical systems. 
The dissipative part, on other hand, is proportional to the variance of the order parameter and thus diverges at the critical point. 

{\bf \emph{Driven-dissipative systems - }}
We consider a system described by a set of bosonic modes $a_i$ evolving according to the master equation
\begin{equation}\label{M}
\partial_t \rho = -i \bigg[H_0 + i \sum\limits_i\mathcal{E}_i(a_i^\dagger - a_i),\rho\bigg]+ \sum\limits_i 2\kappa_i \bigg( a_i \rho a_i^\dagger - \frac{1}{2} \{a_i^\dagger a_i, \rho\}\bigg),    
\end{equation}
where $H_0$ is the Hamiltonian, $\mathcal{E}_i$ are  external pumps  and $\kappa_i$ are the loss rates for each mode [see Fig.~\ref{fig:diagrams}(a)]. 
We work in phase space by defining the  Husimi  function  $Q(\mu,\bar{\mu}) = \frac{1}{\pi} \langle \mu | \rho | \mu \rangle$, where $|\mu\rangle = \bigotimes_{i} |\mu_i\rangle$ are coherent states and $\bar{\mu}$ denotes complex conjugation. 
The master Eq.~(\ref{M}) is then converted into a Quantum Fokker-Planck (QFP) equation \cite{gardiner2004quantum}
\begin{equation}\label{QFP}
    \partial_t Q = \mathcal{U}(Q) + \sum\limits_i \bigg( \partial_{\mu_i} J_i(Q) + \partial_{\bar{\mu}_i} \bar{J}_i(Q)\bigg),
\end{equation}
where $\mathcal{U}(Q)$ is a differential operator related to the unitary part (see~\cite{SupMat} for examples) and $J_i(Q) = \kappa_i (\mu_i Q + \partial_{\bar{\mu}_i} Q)$ are irreversible quasiprobability currents associated with the photon loss dissipators. 

As our basic entropic quantifier, we use the Shannon entropy of $Q$, known as  Wehrl's entropy~\cite{Wehrl1978GeneralEntropy},
\begin{equation}\label{Wehrl}
    S(Q) = - \int d^2\mu \; Q\ln Q.
\end{equation}
This quantity can be attributed an operational interpretation by viewing $Q(\mu, \bar{\mu})$ as the probability distribution for the outcomes of a heterodyne measurement. 
$S(Q)$ then quantifies the entropy of the system convoluted with the additional noise introduced by the heterodyning~\cite{Wodkiewicz1984,General1995}. 
As a consequence, $S(Q) \geq S(\rho)$, with both  converging in the semi-classical limit.

Next, we differentiate Eq.~(\ref{Wehrl}) with respect to time and use Eq.~(\ref{QFP}). 
Employing a standard procedure developed for classical systems~\cite{Seifert2012}, we can separate $dS/dt$ as in Eq.~(\ref{entropy_rate}), with an entropy flux rate  given by 
\begin{equation}\label{Phi}    
    \Phi = \sum\limits_i 2\kappa_i \langle a_i^\dagger a_i \rangle,
\end{equation}
and an entropy production rate
\begin{equation}\label{Pi}    
    \Pi  = - \int d^2\mu \; \mathcal{U}(Q) \ln Q+ \sum\limits_i \frac{2}{\kappa_i} \int d^2\mu \frac{|J_i(Q)|^2}{Q},
\end{equation}
The entropy flux is seen to be always non-negative, which is a consequence of the fact that the dissipator is at zero temperature, so that  entropy cannot flow from the bath to the system, only the other way around. 
As for $\Pi$ in Eq.~(\ref{Pi}), the last term is the typical dissipative contribution, related to the photon loss channels and also found in~\cite{Santos2017WignerRate}.
The extension to a finite temperature dissipator is straightforward and requires only a small modification of the currents $J_i$~\cite{Santos2017WignerRate}. 
The new feature in Eq.~(\ref{Pi}) is the first term, which is related to the unitary contribution $\mathcal{U}(Q)$.
Unlike the von Neumann entropy, the unitary dynamics can affect the Wehrl entropy.
This is due to the fact that the unitary  dynamics can already lead to diffusion-like terms in the Fokker-Planck Eq.~(\ref{QFP}), as discussed e.g. in  Ref.~\cite{Altland2012}.

{\bf \emph{Thermodynamic limit - }} 
The results in Eqs.~(\ref{Phi})-(\ref{Pi}) hold for a generic master equation of the form~(\ref{M}), irrespective of whether or not the system is critical. 
We now reach the key part of our paper, which is to specialize the previous results to the scenario of driven-dissipative critical systems. 
The first ingredient that is needed is the notion of a thermodynamic limit.
For driven-dissipative systems, criticality emerges when the pump(s) $\mathcal{E}_i$ become sufficiently large. 
It is therefore convenient to parametrize 
$\mathcal{E}_i = \epsilon_i \sqrt{N}$ and define the thermodynamic limit as $N \to \infty$, with $\epsilon_i$ finite. 

In these driven systems  $\langle a_i\rangle$ always scale proportionally to the $\mathcal{E}_i$, so that we can also define $\langle a_i \rangle = \alpha_i \sqrt{N}$, where the $\alpha_i$ are finite and represent the order parameters of the system. 
This combination of scalings imply that at the mean-field level ($a_i \to \langle a_i \rangle$) the pump term $\mathcal{E}_i (a_i^\dagger - a_i)$ in~(\ref{M}) will be  $O(N)$; i.e., extensive. 
We shall henceforth assume that the parameters in the model are such that this is also true for $H_0$ in Eq.~(\ref{M}) (see below for examples).

Introducing displaced operators $\delta a_i = a_i - \alpha_i \sqrt{N}$, the entropy flux~(\ref{Phi}) is  naturally split as 
\begin{equation}\label{Phi_sep}
    \Phi = \Phi_\text{ext} + \Phi_{q} = N\sum\limits_i 2\kappa_i  |\alpha_i|^2 + \sum\limits_i 2 \kappa_i \langle \delta a_i^\dagger \delta a_i \rangle. 
\end{equation}
The first term is extensive in $N$ and depends solely on the mean-field values $|\alpha_i|$. 
It is thus independent of fluctuations.
The second term, on the other hand, is intensive in $N$. In fact, it is proportional to the  variance of the order parameter $\langle \delta a_i^\dagger \delta a_i \rangle$ (the susceptibility) and thus captures the contributions from quantum fluctuations.  

We can also arrive at a similar splitting for the entropy production~(\ref{Pi}). 
Defining displaced phase-space variables $\nu_i = \mu_i - \alpha_i \sqrt{N}$, the currents $J_i$ in the QFP Eq.~(\ref{QFP}) are split as $J_i =\sqrt{N} \kappa_i \alpha_i  Q + J_i^{\nu}(Q)$, where $J_i^{\nu} = \kappa_i (\nu_i Q + \partial_{\nu_i^*} Q)$.
Substituting  in~(\ref{Pi}) then yields
\begin{IEEEeqnarray}{rCl}\label{Pi_sep}
    \Pi &=& \Pi_\text{ext} + \Pi_u+ \Pi_d \\[0.2cm]
    &=&  N\sum\limits_i 2\kappa_i  |\alpha_i|^2 
    - \int d^2\nu \; \mathcal{U}(Q) \ln Q
    + \sum\limits_i \frac{2}{\kappa_i}  \int d^2\nu\; \frac{|J_i^{\nu}(Q)|^2}{Q}. 
    \nonumber
\end{IEEEeqnarray}
This is the main  result in this paper. 
It offers a splitting of the total entropy production rate into three contributions with distinct physical interpretations.
The first, $\Pi_\text{ext}$, is  extensive and depends solely on the mean-field values $\alpha_i$. 
It therefore corresponds to a fully deterministic contribution, independent of fluctuations. 
Comparing with Eq.~(\ref{Phi_sep}), we see that 
\begin{equation}\label{balance_ext}
\Pi_\text{ext} = \Phi_\text{ext},
\end{equation}
a balance which holds irrespective of whether the system is in the NESS. 
Hence, this contribution does not affect the system entropy: At the mean-field level, all entropy produced flows to the environment. 

The second and third terms in Eq.~(\ref{Pi_sep}) represent, respectively, the unitary and dissipative contributions to $\Pi$. 
These two terms  account for the  contributions to the entropy production stemming from quantum fluctuations.
This becomes more evident in the NESS ($dS/dt = 0$), where combining Eqs.~(\ref{entropy_rate}) and~(\ref{balance_ext}) leads to 
\begin{equation}\label{balance_q}
    \Pi_u + \Pi_d = \Phi_q.
\end{equation}
The two terms $\Pi_u$ and $\Pi_d$ therefore represent two sources for the quantum entropy $\Phi_q$ in Eq.~(\ref{Phi_sep}). 
We also note in passing that while $\Pi_d \geq 0$, the same is not necessarily true for $\Pi_u$, although this turns out to be the case in the examples treated below.

%
%
%
%

\begin{figure}
    \centering
    \includegraphics[width=0.22\textwidth]{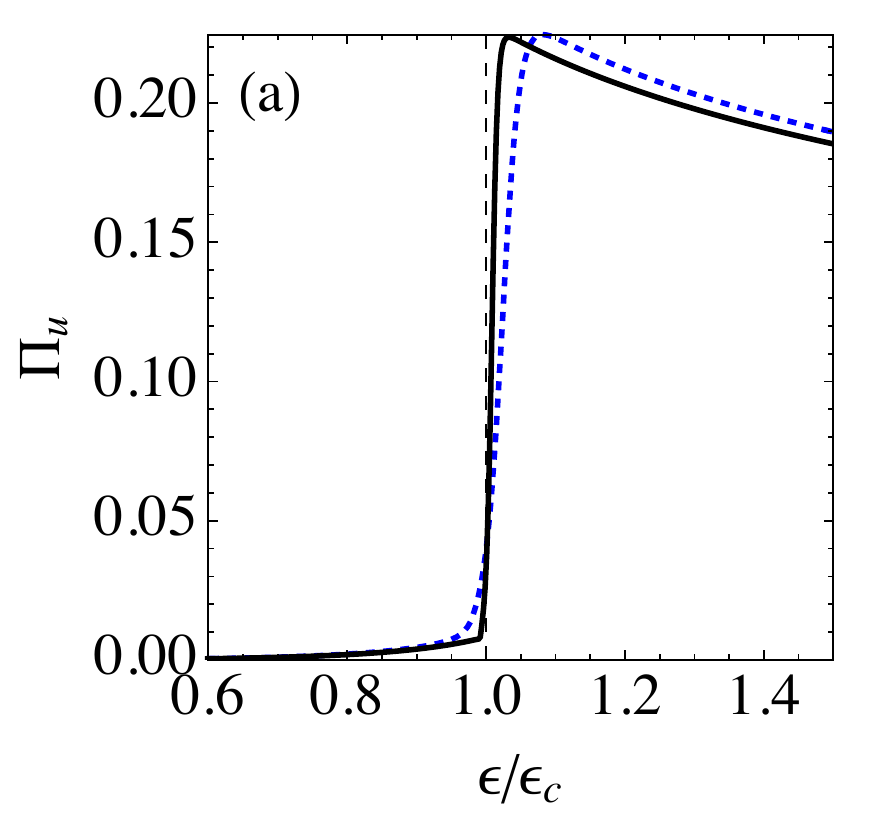}\quad
    \includegraphics[width=0.22\textwidth]{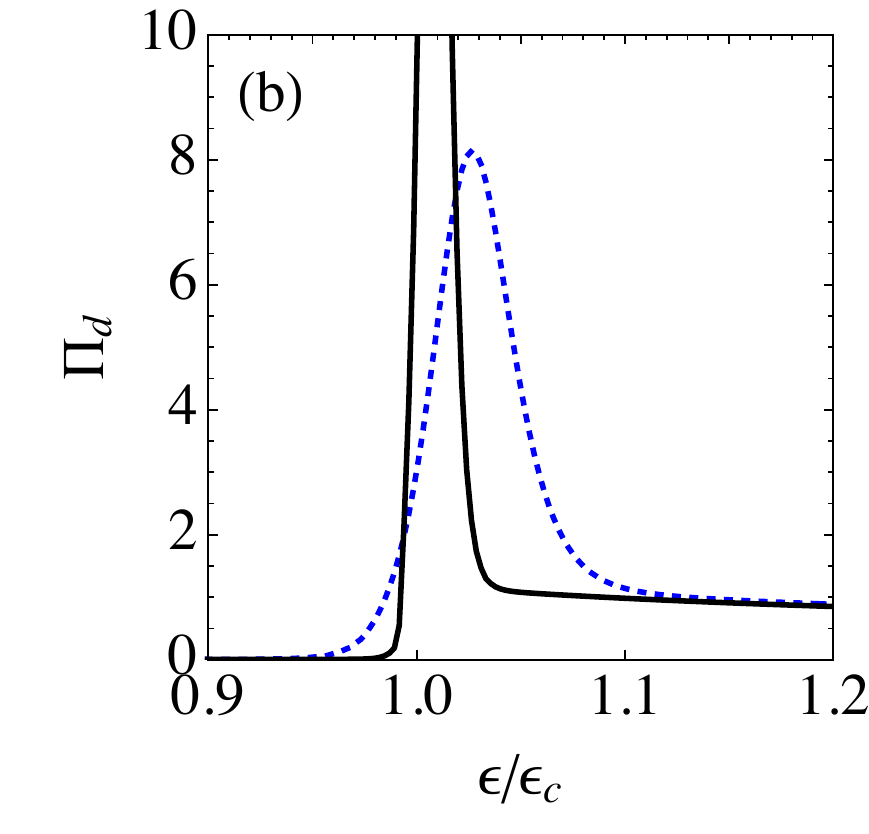}\\[0.2cm]
    \includegraphics[width=0.22\textwidth]{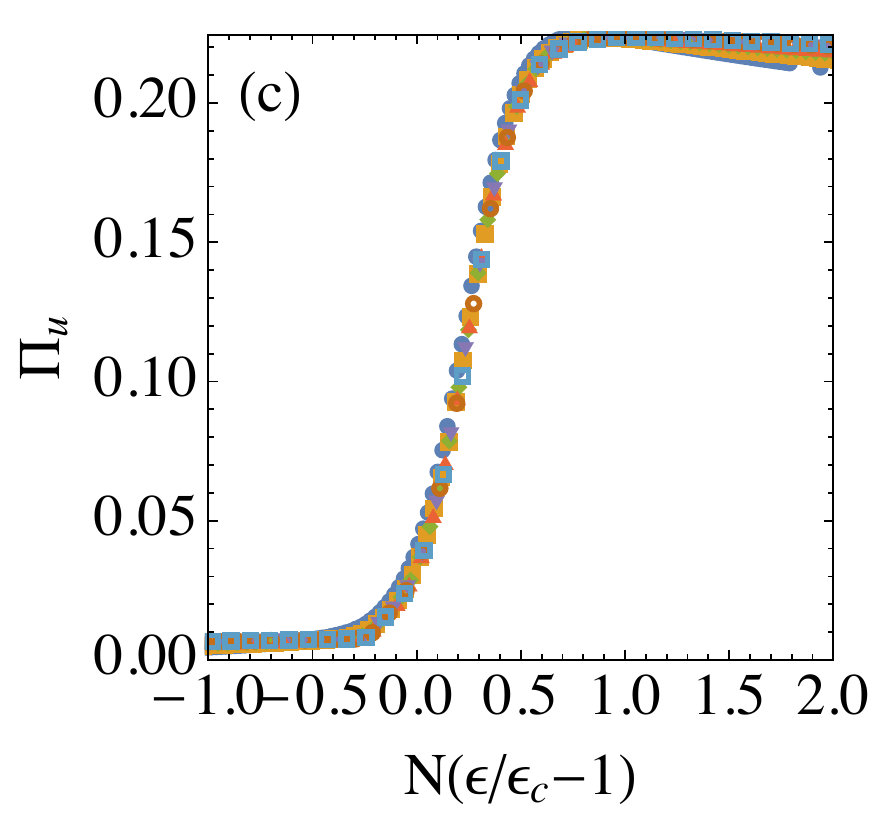}\quad
    \includegraphics[width=0.22\textwidth]{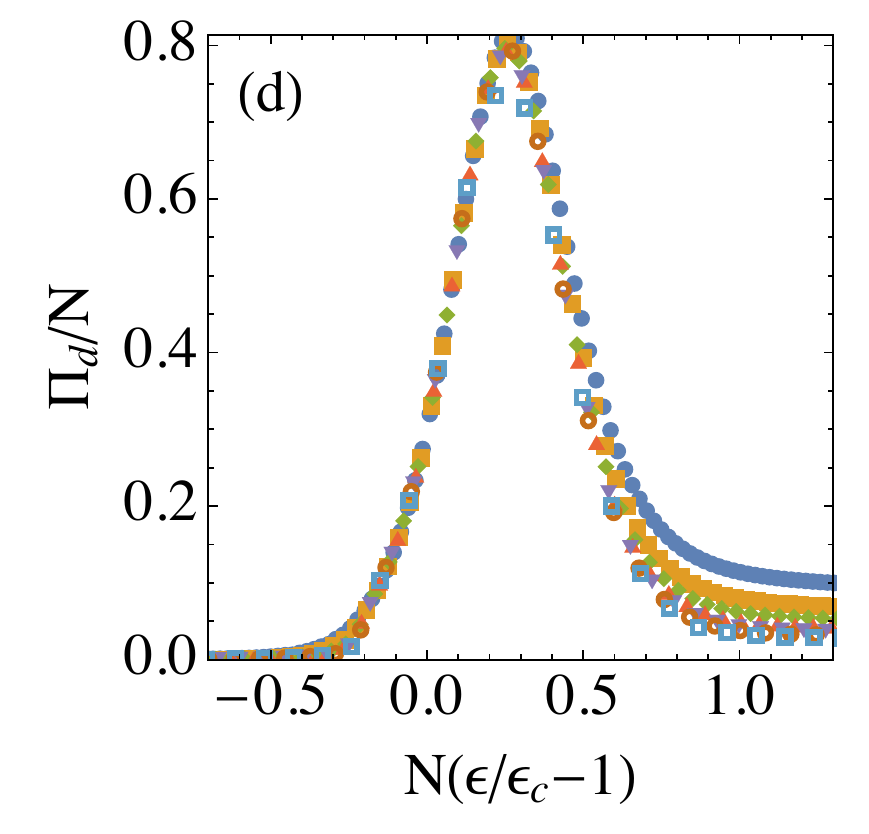}
    \caption{Entropy production in the discontinuous transition of the Kerr bistability model [Eq.~(\ref{H0_Kerr})]. (a),(b)  unitary and dissipative contributions $\Pi_u$ and $\Pi_d$ for $N = 30$ (black-solid) and $10$ (blue-dashed). 
    (c), (d) Finite-size analysis showing a data collapse of $\Pi_u$ and $\Pi_d/N$~vs.~$N(\epsilon/\epsilon_c-1)$ for multiple values of $N$ (from 10 to 40 in steps of 5). 
    The critical behavior of $\Pi_u$ matches that of the classical entropy production. $\Pi_d$, on the other hand, behaves similarly to $\langle \delta a^\dagger \delta a \rangle$ and thus diverges at the critical point. 
    Other parameters were $\kappa = 1/2$, $\Delta = -2$ and $u = 1$.
    }
    \label{fig:kerr}
\end{figure}

{\bf \emph{Kerr bistability - }} To illustrate how the different contributions to the entropy production in Eq.~(\ref{Pi_sep}) behave across a dissipative transition, we now apply our formalism to two prototypical models. The first is the Kerr bistability model~\cite{drummond1980quantum,Carusotto2013,Casteels2017}, described by Eq.~(\ref{M}) with a the single mode $a$ and  Hamiltonian 
\begin{equation}\label{H0_Kerr}
    H_0 = \Delta a^\dagger a + \frac{u}{2N} a^\dagger a^\dagger a a,
\end{equation}
where $\Delta$ is the detuning and $u$ is the non-linearity strength. 
This model has a discontinuous transition. 

The NESS of this model and the terms in Eq.~(\ref{Pi_sep}) were computed using numerically exact methods. 
Details on the numerical calculations are provided  in the Supplemental Material~\cite{SupMat} and the main results are shown in Fig.~\ref{fig:kerr}. 
In Figs.~\ref{fig:kerr}(a) and (b) we plot $\Pi_u$ and $\Pi_d$ for different sizes $N$. As can be seen, $\Pi_u$ has a discontinuity  at the critical point when $N\to \infty$. Conversely, $\Pi_d$ diverges. 
The critical behavior in the thermodynamic limit ($N\to \infty$) can be better understood by performing a finite size analysis (Figs.~\ref{fig:kerr}(c) and (d)), where we plot $\Pi_u$ and $\Pi_d/N$~vs.~$N(\epsilon/\epsilon_c-1)$ for multiple values of $N$.
Surprisingly, we find that the behavior of $\Pi_u$ matches exactly that of the classical entropy production in a discontinuous transition~\cite{Noa2019,Zhang2016,Herpich2019} (see~\cite{SupMat} for more information).
We also see from Fig.~\ref{fig:kerr} that $\Pi_u$ is negligible compared to $\Pi_d$.
As a consequence, in view of Eq.~(\ref{balance_q}) the dissipative contribution $\Pi_d$ will behave like the variance of the order parameter $\langle \delta a^\dagger \delta a\rangle$, which diverges at the critical point. 
This is clearly visible in Fig.~\ref{fig:kerr}(d), which plots $\Pi_d/N$.

%
%
%
%

\begin{figure}
    \centering
    \includegraphics[width=0.22\textwidth]{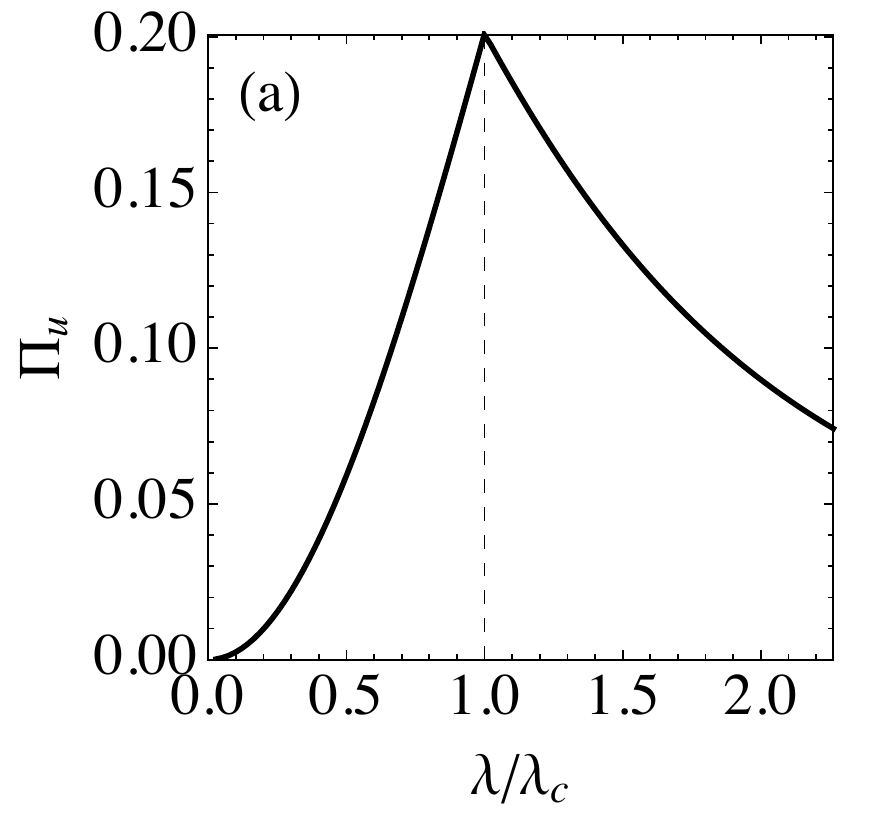}\quad
    \includegraphics[width=0.22\textwidth]{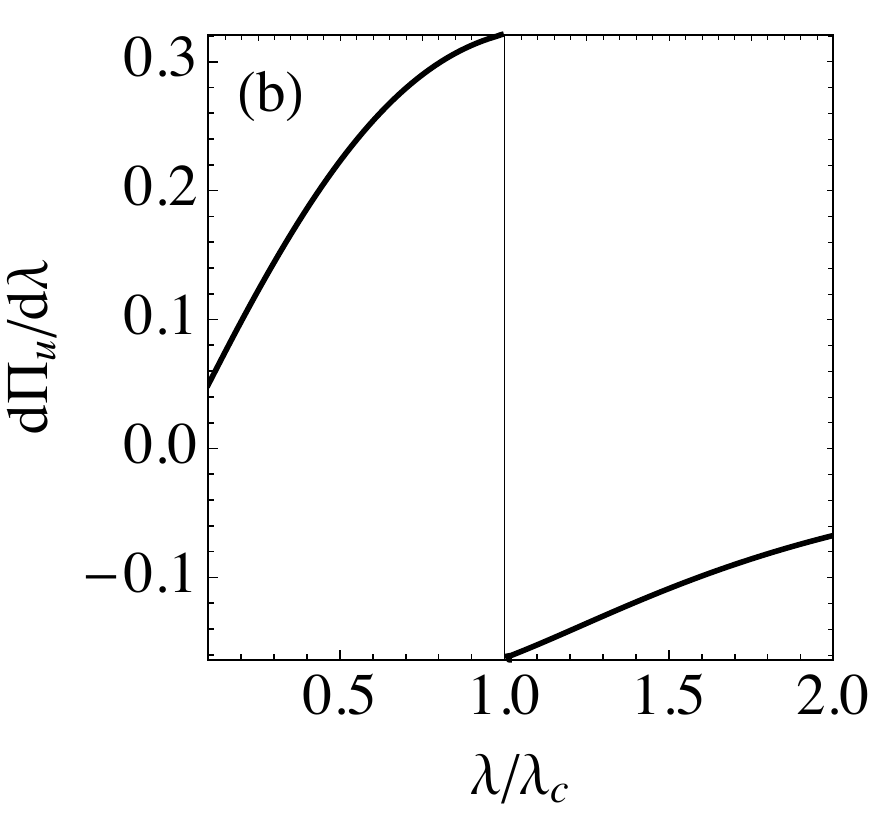}
    \\[0.2cm]
    \includegraphics[width=0.22\textwidth]{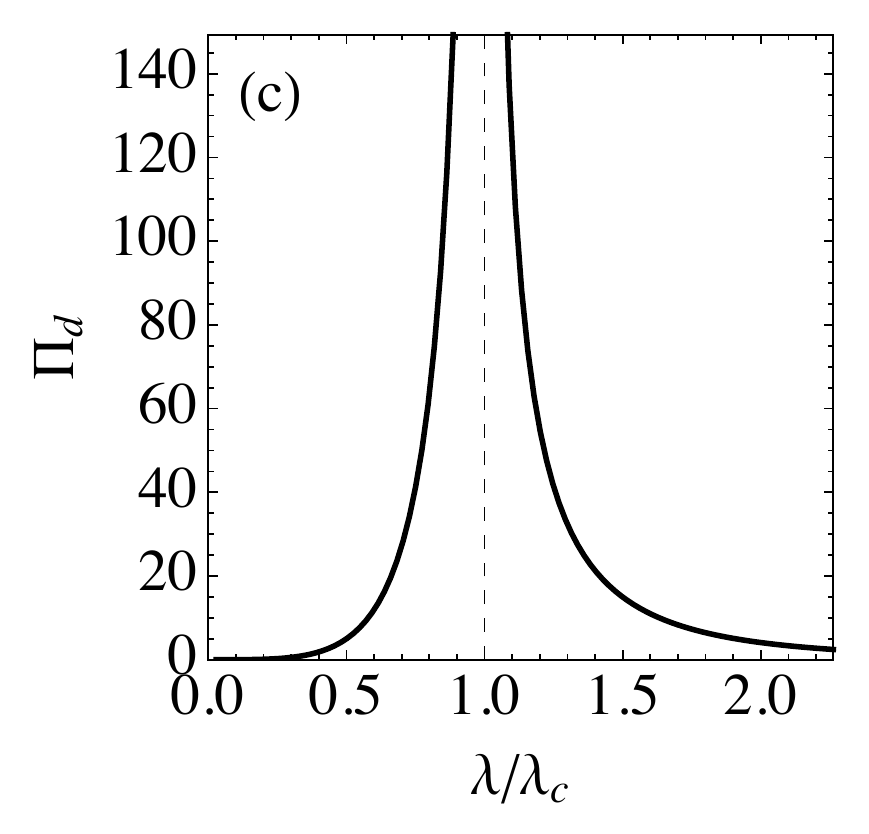}\quad
    \includegraphics[width=0.22\textwidth]{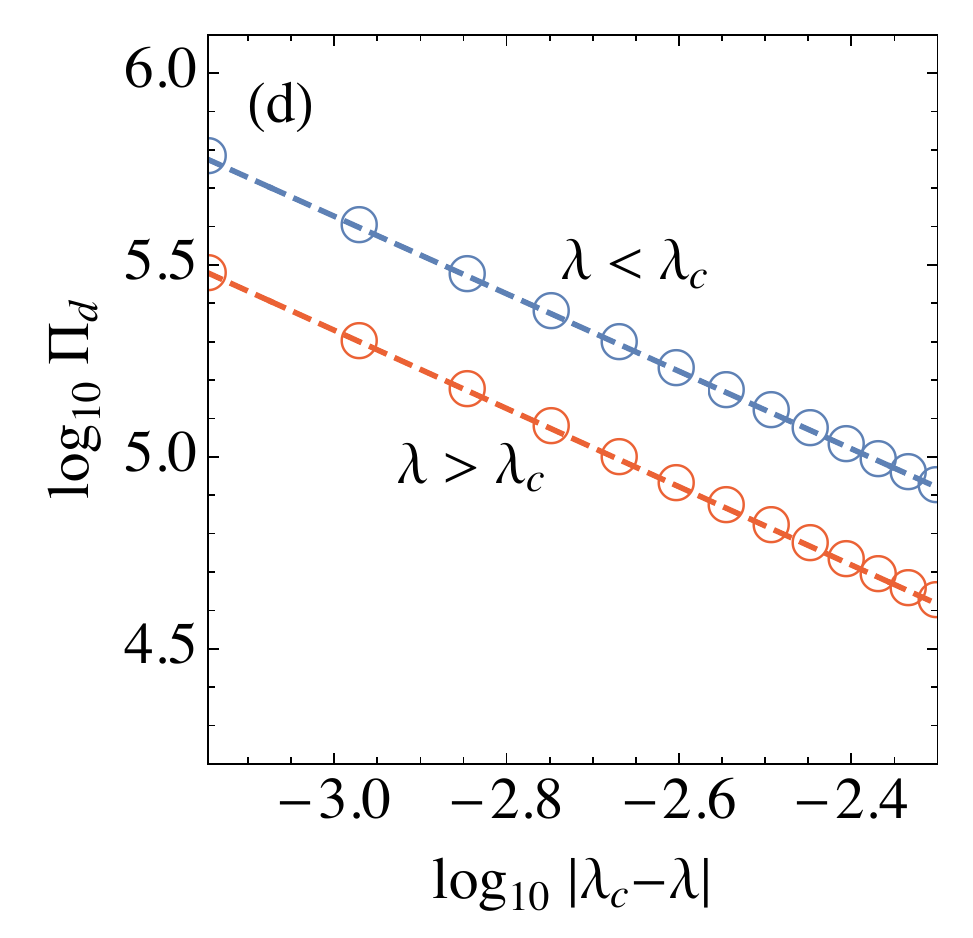}
    \caption{Entropy production in the continuous transition of the driven-dissipative Dicke model [Eq.~(\ref{H0_Dicke})].
    (a), (b) $\Pi_u$ and  $d\Pi_u/d\lambda$~vs.~$\lambda$. This part of the entropy production is continuous, but has a kink (discontinuous first derivative) at the critical point $\lambda_c = \sqrt{\omega_0 (\kappa^2 + \omega^2)/\omega}$. 
    (c) $\Pi_d$~vs.~$\lambda$ showing a divergence at $\lambda_c$. 
    (d) $\log_{10}\Pi_d$~vs.~$\log_{10} |\lambda_c - \lambda|$ at the vicinity of $\lambda_c$. The points correspond to simulations, whereas the straight lines are fits with slope $-1$,  showing that $\Pi_d$ diverges as in Eq.~(\ref{Dicke_divergence}). 
    Other parameters were $\omega_0 = 0.005$, $\omega = 0.01$ and $\kappa = 1$.
    }
    \label{fig:dicke}
\end{figure}

{\bf \emph{Driven-dissipative Dicke model - }} 
The second model we study is the driven-dissipative Dicke model~\cite{Baumann2010,Landig2015}.
It is described by  Eq.~(\ref{M}) with a mode $a$, subject to photon loss dissipation $\kappa$, as well as a macrospin of size $J= N/2$. The Hamiltonian is 
\begin{equation}\label{H0_Dicke}
    H_0 = \omega_0 J_z + \omega a^\dagger a + \frac{2\lambda}{\sqrt{N}} (a + a^\dagger)J_x,
\end{equation}
where $J_i$ are macrospin operators.
This model does not need a  drive $\mathcal{E}$ since the last term can already be interpreted as a kind of ``operator valued pump'' (as it is linear in $a+a^\dagger$).
In fact, this is precisely how this model was experimentally implemented in a cold-atom setup~\cite{Baumann2010}. 
The model can also be pictured as  purely bosonic by introducing an additional mode $b$ according to the Holstein-Primakoff map $J_z = b^\dagger b - N/2$ and $J_- = \sqrt{N - b^\dagger b} b$.
It hence falls under the category of Eq.~(\ref{M}), with two modes $a$ and $b$. 

Since this is a two-mode model, numerically exact results are more difficult. 
Instead, we follow Ref.~\cite{Baumann2010,Landig2015,Brunelli2018ExperimentalSystems} and consider a Gaussianization of the model valid in the limit of $N$ large. 
Details are provided in~\cite{SupMat} and the results are shown in Fig.~\ref{fig:dicke}. 
Once again, the unitary part $\Pi_u$ of the entropy production (Figs.~\ref{fig:dicke}(a) and (b)) is found to behave like the mean-field predictions for classical transitions~\cite{Tome2012,Shim2016,Crochik2005,Zhang2016,Noa2019,Herpich2018,Herpich2019}.
It is continuous and finite, but presents a kink (the first derivative is discontinuous) at the critical point $\lambda_c = \sqrt{\omega_0 (\kappa^2 + \omega^2)/\omega}$.


The dissipative part $\Pi_d$, on the other hand, diverges at $\lambda_c$. 
This was in fact already shown experimentally in Ref.~\cite{Brunelli2018ExperimentalSystems}.
In fact, the behavior of $\Pi_d$ at the vicinity of $\lambda_c$ is of the form 
\begin{equation}\label{Dicke_divergence}
    \Pi_q \sim \frac{1}{|\lambda_c - \lambda|},
\end{equation}
as confirmed by the analysis in Fig.~\ref{fig:dicke}(d). 
Similarly to the Kerr model, $\Pi_u$ is much smaller than $\Pi_d$ so that the latter essentially coincides with $2\kappa \langle \delta a^\dagger \delta a \rangle$ in the NESS (c.f. Eq.~(\ref{balance_q})).
The divergence in~(\ref{Dicke_divergence}) thus mimics the behavior of  $\langle \delta a^\dagger \delta a \rangle$.

%
%
%
%

{\bf \emph{Discussion - }} Understanding the behavior of the entropy production across a non-equilibrium transition is both a timely and important question, specially concerning driven-dissipative quantum models, which have found renewed interest in recent years. 
The reason why this problem was not studied before, however, was because there were \emph{no} theoretical frameworks available for computing the entropy production for the zero-temperature dissipation appearing in driven-dissipative models.
This paper provides such a framework.

We then applied our formalism to two widely used models. 
In both cases we found that one contribution $\Pi_u$ behaved qualitatively similar to that of the  entropy production in classical dissipative transitions. 
The other, $\Pi_d$, behaved like a susceptibility, diverging at the critical point. 
Why $\Pi_u$ behaves in this way, remains an open question. 
Driven-dissipative systems have one fundamental difference when compared to classical systems. 
In the latter,  energy input and output both take place incoherently, through the transition rates in a master equation. 
In driven-dissipative systems, on the other hand, the energy output is incoherent (Lindblad-like) but the input is coherent (the pump). 
A classical analog of this is an electrical circuit coupled to an external battery $\mathcal{E}$.
For instance, the entropy production in a simple RL circuit  is $\Pi_\text{RL} = \mathcal{E}^2/RT$~\cite{Landauer1975} where $R$ is the resistance and $T$ is the temperature. 
If we consider an empty cavity with a single mode $a$ and $H_0 = 0$, Eq.~(\ref{Pi}) predicts $\Pi_\text{cavity} =2 \mathcal{E}^2/\kappa$. 
Notwithstanding the similarity between the two results, one must bear in mind that the RL circuit still contains incoherent energy input. 
Indeed, $\Pi_\text{RL}$ diverges as $T\to 0$. 
The cavity, on the other hand, relies solely on vacuum fluctuations. 
This interplay between thermal and quantum fluctuations highlights the need for extending the present analysis to additional models of driven-dissipative transitions. In particular, it would be valuable to explore models which can be tuned between classical ($T\neq0$) and quantum ($T=0$) transitions.


\emph{Acknowledgments} - The authors acknowledge fruitful discussions with M. Paternostro, T. Donner, J. P. Santos, W. Nunes and L. C. C\'eleri. 
The authors acknowledge Raam Uzdin for fruitful discussion, as well as for coining the term ``zero temperature catastrophe''. 
GTL acknowledges the financial support from the S\~ao Paulo Research Foundation (FAPESP) under grants 2018/12813-0, 2017/50304-7 and 2017/07973-5. BOG acknowledges the support from the brazilian funding agency CNPq and T. Roberta for drawing fig. (1).

\bibliography{references}

\pagebreak
\widetext

\newpage 
\begin{center}
\vskip0.5cm
{\Large Supplemental Material}
\end{center}

\setcounter{equation}{0}
\setcounter{figure}{0}
\setcounter{table}{0}
\setcounter{page}{1}
\renewcommand{\theequation}{S\arabic{equation}}
\renewcommand{\thefigure}{S\arabic{figure}}

This supplemental material is divided in three parts. 
In Sec.~S1  we provide additional details on the structure of the unitary contribution $\Pi_u$ appearing in Eq.~(\ref{Pi_sep}) of the main text.
Then, in Secs.~S2 and S3, we  provide technical details on the two applications, the Kerr and Dicke models, studied in the main text. 

\section{S1. Properties of $\Pi_u$}

The entropy production rate in Eq.~(\ref{Pi_sep}) of the main text has a term proportional to the unitary dynamics, 
\begin{equation}\label{SM_Piu}
    \Pi_u = - \int d^2\mu \; \mathcal{U}(Q) \ln Q,
\end{equation}
which depends on the differential operator $\mathcal{U}(Q)$, representing the unitary contribution to the QFP Eq.~(\ref{QFP}).
Written in this way, the physics behind this term is not immediately transparent. To shed light on this, we focus here the case of a single mode. 
The Hamiltonian may then always be written in normal order as
\begin{equation}\label{SM_H0_scaling_1}
    H_0 = \sum\limits_{r,s} H_{r s} (a^\dagger)^r a^s, 
\end{equation}
for some coefficients $H_{r s}$.
The thermodynamic limit hypothesis used in the main text is that $H_0$ should be $O(N)$ at the mean-field level ($a_i \to \langle a_i \rangle$). 
This implies that $H_{rs} = h_{rs} N^{1-(r+s)/2}$, where the $h_{rs}$ are independent of $N$. 
For instance, the coefficient multiplying $a^\dagger a^\dagger a a$ should scale as $1/N$ (as in Eq.~(\ref{H0_Kerr})). 
We may thus write~(\ref{SM_H0_scaling_1}) as 
\begin{equation}\label{SM_H0_scaling_2}
    H_0 = N \sum\limits_{r,s} h_{r s} \left(\frac{a^\dagger}{\sqrt{N}}\right)^r \left(\frac{a}{\sqrt{N}}\right)^s .
\end{equation}

The corresponding phase-space contribution $\mathcal{U}(Q)$ can be found using standard correspondence tables~\cite{gardiner2004quantum} and reads
\begin{equation}
    \mathcal{U}(Q) = -i N \sum\limits_{r,s} \frac{h_{rs}}{N^{(r+s)/2}} \bigg\{ \bar{\mu}^r (\mu + \partial_{\bar{\mu}})^s - \mu^s (\bar{\mu} + \partial_\mu)^r \bigg\} Q.
\end{equation}
Normal ordering is convenient as it pushes all derivatives to the right. 
We now change variables to $\nu = \mu - \alpha \sqrt{N}$
and expand the result in a power series in $N$. 
This yields, to leading order
\begin{IEEEeqnarray}{rCl}
\label{SM_UQ_expansion}
    \mathcal{U}(Q) &=& -i \sqrt{N} \sum\limits_{r,s} h_{rs} \alpha^{s-1} \bar{\alpha}^{r-1} \left( s \bar{\alpha} \partial_{\bar{\nu}} - r \alpha \partial_\nu\right) Q \\[0.2cm] \nonumber
    &&-i  \sum\limits_{r,s} h_{rs} \frac{\alpha^{s-2}\bar{\alpha}^{r-2}}{2} \Bigg[ 
      s(s-1) (\bar{\alpha})^2 (2 \nu \partial_{\bar{\nu}} + \partial_{\bar{\nu}}^2) 
    - r(r-1) \alpha^2 (2 \bar{\nu} \partial_\nu + \partial_\nu^2) 
    + 2 r s |\alpha|^2 (\bar{\nu}\partial_{\bar{\nu}} - \nu \partial_\nu) \Bigg] Q \\[0.2cm]
    &&+ \mathcal{O}(1/\sqrt{N}).
    \nonumber
\end{IEEEeqnarray}
The remaining terms in the expansion are at least $O(1/\sqrt{N})$ and thus vanish in the limit $N \to \infty$. 
This expression may be further simplified by introducing the
 constants 
\begin{IEEEeqnarray}{rCl}
    \xi_1 &=& -i \sum\limits_{r,s} h_{rs} \; \alpha^{s-1} \bar{\alpha}^{r} s, \\[0.2cm]
    \xi_2 &=& - i \sum\limits_{r,s} h_{rs} \; \alpha^{s-2} \bar{\alpha}^{r} s(s-1), \\[0.2cm]
    \xi_{11} &=& - i \sum\limits_{r,s} h_{rs}\; \alpha^{s-1} \bar{\alpha}^{r-1} r s. 
\end{IEEEeqnarray}
Then, since $h_{r s} = h_{s r}^*$, we can write~(\ref{SM_UQ_expansion}) as 
\begin{equation}\label{SM_UQ_expansion2}
    \mathcal{U}(Q) = \sqrt{N}\bigg(\xi_1 \partial_{\bar{\nu}} + \bar{\xi_1} \partial_\nu\bigg) Q + \frac{1}{2} \bigg[ \xi_2 (2 \nu \partial_{\bar{\nu}} + \partial_{\bar{\nu}}^2)+ \bar{\xi}_2 (2 \bar{\nu} \partial_\nu + \partial_\nu^2) + 2 \xi_{11}(\bar{\nu} \partial_{\bar{\nu}} - \nu \partial_\nu) \bigg] Q + O(1/\sqrt{N}).
\end{equation}
This is the leading order contributions of the unitary dynamics to the Fokker-Planck equation. 
The important point to notice is the existence of \emph{diffusive} terms (proportional to the second derivative $\partial_\nu^2$ and $\partial_{\bar{\nu}}^2$). 
This is a known feature of the Husimi function.

We now plug Eq.~(\ref{SM_UQ_expansion2}) into Eq.~(\ref{SM_Piu}).
Integrating by parts multiple times and using the fact that the Husimi function always vanishes at infinity, we find that the only surviving terms are 
\begin{equation}\label{SM_Piu_leading_order}
     \Pi_u = \frac{1}{2} \int\frac{d^2\nu}{Q} \bigg[\xi_2 (\partial_{\bar{\nu}} Q)^2 + \bar{\xi}_2 (\partial_\nu Q)^2 \bigg],
\end{equation}
which provides the leading order contribution to $\Pi_u$. In the limit $N\to \infty$ this is the only contribution which survives. 


\section{S2. Kerr bistability}

In this section we provide additional details on the solution methods used to study the entropy production in the Kerr model [Eq.~(\ref{H0_Kerr}) of the main text]. 
The NESS of this model can be found analytically using the generalized P function~\cite{drummond1980quantum}. 
This includes all moments of the form $\langle (a^\dagger)^r a^s \rangle$, as well as the Wigner function~\cite{Kheruntsyan1999}. 
While the Husimi function can in principle be found numerically from the Wigner function, we have found that this is quite numerically unstable due to the highly irregular nature of the latter.
Instead, it is  easier to simply find the steady-state density matrix $\rho$ numerically using standard vectorization techniques (as done e.g. in Ref.~\cite{Casteels2017}).

\subsection{Numerical procedure}

The numerical calculations were performed as follows. 
We define the Liouvillian corresponding to the master equation~(\ref{M}) as 
\begin{equation}\label{SM_Kerr_Liouvillian}
    \mathcal{L}(\rho) = -i \bigg[H_0 + i \epsilon \sqrt{N}(a^\dagger - a),\rho\bigg]+  2\kappa \bigg( a \rho a^\dagger - \frac{1}{2} \{a^\dagger a, \rho\}\bigg).
\end{equation}
The steady-state equation, 
\begin{equation}
\mathcal{L}(\rho) = 0,
\end{equation}
is then interpreted as an eigenvalue/eigenvector equation: $\rho$ is the eigenvector of $\mathcal{L}$ with eigenvalue $0$. 
To carry out the calculation, we decompose $\rho$ in the Fock basis, using a sufficiently large number of states $n_\text{max}$ to ensure convergence. 

From $\rho$ we then compute the Husimi function and the corresponding integrals numerically using standard integration techniques. 
The Husimi function is obtained by constructing approximate coherent states 
\[
|\mu\rangle = e^{-|\mu|^2/2} \sum\limits_{n=0}^{n_\text{max}} \frac{\mu^n}{\sqrt{n!}} |n\rangle.  
\]
A grid of the Husimi function $Q(\mu,\bar{\mu})$ can then be  built to be subsequently integrated numerically. 
Derivatives of $Q$ do not need to be computed using finite differences. Instead, one may notice that, for instance,
\begin{equation}
    \partial_{\bar{\mu}} Q = - \mu Q + \frac{1}{\pi} \langle \mu | a \rho | \mu\rangle,
\end{equation}
with similar expressions for other derivatives. 
Finally, convergence of the numerical integration can be verified by computing moments $\langle (a^\dagger)^r a^s \rangle$ of arbitrary order from the Husimi function and comparing with the exact results of Ref.~\cite{drummond1980quantum}. 

\subsection{Bistable behavior}

For fixed $\kappa$, $U$ and $\Delta <0$, the NESS of Eq.~(\ref{SM_Kerr_Liouvillian}) presents a discontinuous transition at a certain critical value $\epsilon_c$. 
This transition is related to a bistable behavior of the model at the mean-field level. 
For finite $N$ the steady-state of~(\ref{SM_Kerr_Liouvillian}) is unique~\cite{drummond1980quantum}. 
However, as  shown recently in Ref.~\cite{Casteels2017}, in the limit $N\to \infty$ the Liouvillian gap between the steady-state and the first excited state closes asymptotically in the region between
\begin{equation}
    \epsilon_\pm = \sqrt{n_\pm \big[\kappa^2 + (\Delta + n_\pm u)^2\big]}, \qquad
    n_\pm = \frac{-2 \Delta \pm \sqrt{\Delta^2 - 3 \kappa^2}}{3 u}. 
\end{equation}
From a numerical point of view, however, this causes no interference since all computations are done for finite $N$, where the NESS is unique.

\subsection{Unitary contribution to the Quantum Fokker-Planck equation}

The unitary contribution $\mathcal{U}(Q)$ appearing in Eq.~(4) of the main text can be obtained using standard correspondence tables~\cite{gardiner2004quantum} and reads
\begin{equation}
    \mathcal{U}(Q) = (i  \mu \Delta -\mathcal{E}) \partial_\mu Q - (i  \bar{\mu} \Delta+\mathcal{E})\partial_{\bar{\mu}}Q
    + \frac{i u}{2N} \bigg\{ 2 |\mu|^2 \big(\mu \partial_\mu Q - \bar{\mu} \partial_{\bar{\mu}}Q\big) + \mu^2 \partial_\mu^2 Q - \bar{\mu}^2\partial_{\bar{\mu}}^2 Q\bigg\}.
\end{equation}
When plugged into Eq.~(\ref{SM_Piu}), the terms proportional to $\Delta$ and $\mathcal{E}$ vanish. 
The only surviving terms are
\begin{equation} 
\Pi_u = \frac{i u}{2N} \int \frac{d^2\mu}{Q} \bigg[ \mu^2 (\partial_\mu Q)^2 - \bar{\mu}^2 (\partial_{\bar{\mu}}Q)^2\bigg].
\end{equation}
Substituting $\mu = \alpha \sqrt{N} + \nu$ yields a leading contribution of $O(1)$ which, of course, is the same as that which would be obtained using Eq.~(\ref{SM_Piu_leading_order}) with $r = s = 2$. 



\section{S3. Driven-dissipative Dicke model}

Here we describe the calculations for the driven-dissipative Dicke model [Eq.~(\ref{H0_Dicke}) of the main text]. We consider only a single source of drive and dissipation $(\mathcal{E},\kappa)$ acting on the optical cavity mode $a$. 
The full master equation is then 
\begin{equation}
    \frac{d \rho}{d t} = -i [H,\rho] + 2 \kappa \bigg[a \rho a^\dagger - \frac{1}{2} \{a^\dagger a, \rho\} \bigg],
\end{equation}
with 
\begin{equation}\label{Dicke_H}
    H = \omega_0 J_z + \omega a^\dagger a + \frac{2\lambda}{\sqrt{N}} (a+a^\dagger)J_x 
\end{equation}
Since this system involves two modes, direct solution by vectorization becomes computationally too costly. 
Instead, we tackle the problem using Gaussianization. 
The calculations are done in detail  in Refs.~\cite{Baumann2010,Landig2015,Brunelli2018ExperimentalSystems}. Here we simply cite the main results and  adapt the notation to our present interests. 

\subsection{Mean-field solution}

We start by looking at the mean-field level by introducing $\langle a \rangle = \alpha \sqrt{N}$, $\langle J_- \rangle = \beta N$ and $\langle J_z \rangle = w N$. For large $N$ we then get 
\begin{IEEEeqnarray}{rCl}
\label{Dicke_MFeqs_alpha}
    \frac{d \alpha}{d t} &=& - (\kappa + i \omega) \alpha - i \lambda (\beta + \beta^*) ,    \\[0.2cm]
\label{Dicke_MFeqs_beta}    
    \frac{d \beta}{d t} &=& - i \omega_0 \beta + 2 i \lambda (\alpha + \bar{\alpha}) w ,   \\[0.2cm]
\label{Dicke_MFeqs_w}    
    \frac{dw}{d t} &=& i \lambda (\alpha + \bar{\alpha}) (\beta - \beta^*),
\end{IEEEeqnarray}
which are independent of $N$, as expected. 
Angular momentum conservation also imposes $w^2 + |\beta|^2 = 1/4$, which leads to two choices, $w = \pm \frac{1}{2} \sqrt{1- 4 \beta^2}$. 

At the steady-state this implies that $\beta^* = \beta$,
\begin{equation}
    \alpha = 
    - \frac{2 i \lambda \beta}{\kappa+ i \omega},
\end{equation}
and 
\begin{equation}\label{Dicke_mean_field_beta}
    -\beta 
    \sqrt{1 - 4 \beta^2} = \pm\frac{\lambda_c^2}{\lambda^2} \beta,
\end{equation}
where $\lambda_c = \frac{1}{2} \sqrt{\frac{\omega_0}{\omega} (\kappa^2 + \omega^2)}$ is the critical interaction in the absence of any external drives. 
The $\pm$ sign in Eq.~(\ref{Dicke_mean_field_beta}) stems from the two choices $w = \pm \frac{1}{2} \sqrt{1- 4 \beta^2}$ respectively.
The minus solution in Eq.~(\ref{Dicke_mean_field_beta})  always yields the trivial result $\beta = 0$. 
The plus solution, on the other hand, can be non-trivial when $\lambda > \lambda_c$. 
For this reason, we henceforth 
focus on the solution of 
\begin{equation}
    \beta \sqrt{1 - 4\beta^2} = \frac{\lambda_c^2}{\lambda^2} \beta,
\end{equation}
which yields either $\beta = 0$ or $\beta \in [0,1/2]$. 
Moreover, this solution corresponds to $w = -\frac{1}{2} \sqrt{1- 4 \beta^2}$, so that the spin is pointing downwards.

\subsection*{Holstein-Primakoff expansion}

Next we introduce a Holstein-Primakoff expansion 
\begin{IEEEeqnarray}{rCl}
\label{Dicke_HP_Jz}
    J_z &=& b^\dagger b - \frac{N}{2},  \\[0.2cm]
\label{Dicke_HP_Jm}
    J_- &=& \sqrt{N - b^\dagger b} \; b.
\end{IEEEeqnarray}
and expand
\begin{equation}\label{Dicke_HP_ab_expansion}
    a = \alpha \sqrt{N} + \delta a, 
    \qquad
    b = \tilde{\beta} \sqrt{N} + \delta b,
\end{equation}
for $\alpha$ and $\tilde{\beta}$ independent of $N$.
The constant $\tilde{\beta}$ can be related with $\beta = \langle J_- \rangle/N$ by expanding Eq.~(\ref{Dicke_HP_Jm}) in a power series in $1/N$, resulting in
\begin{equation}
    \tilde{\beta}\sqrt{1-\tilde{\beta}^2}  = \beta,
\end{equation}
which has two solutions
\begin{equation}
    \tilde{\beta}_\pm = \sqrt{\frac{1 \pm \sqrt{1 - 4 \beta^2}}{2}}.
\end{equation}
Which solution to choose is fixed by imposing that Eq.~(\ref{Dicke_HP_Jz}) should also comply with $\langle J_z \rangle = w\sqrt{N}$ and $w = -\frac{1}{2} \sqrt{1- 4 \beta^2}$. 
This fixes $\tilde{\beta}_-$ as the appropriate choice. 
It is also useful to note that $\tilde{\beta}_+^2 + \tilde{\beta}_-^2 = 1$ and $\tilde{\beta}_- \tilde{\beta}_+ = \beta$. 

In terms of the expansion~(\ref{Dicke_HP_ab_expansion}) the operator $J_z$ in Eq.~(\ref{Dicke_HP_Jz}) becomes
\begin{equation}\label{Dicke_HP_Jz_expansion}
    J_z = \frac{N}{2} \sqrt{1 - 4\beta^2} + \sqrt{N} \tilde{\beta}_- (\delta b + \delta b^\dagger) + \delta b^\dagger \delta b. 
\end{equation}
We similarly expand $J_-$ in Eq.~(\ref{Dicke_HP_Jm}), leading to  
\begin{equation}\label{Dicke_HP_Jm_expansion}
    J_- = N \beta + \sqrt{N} \tilde{\beta}_+ \bigg[ \delta b - \frac{1}{2} \frac{\tilde{\beta}_-^2}{\tilde{\beta}_+^2} (\delta b + \delta b^\dagger) \bigg] - \frac{\tilde{\beta}_-}{2 \tilde{\beta}_+} \bigg[ \delta b^\dagger \delta b + (\delta b + \delta b^\dagger) \delta b + \frac{\tilde{\beta}_-^2}{4\tilde{\beta}_+^2} (\delta b + \delta b^\dagger)^2 \bigg] + \mathcal{O}(1/\sqrt{N}).
\end{equation}
Substituting Eqs.~(\ref{Dicke_HP_Jz_expansion}) and (\ref{Dicke_HP_Jm_expansion}) into Eq.~(\ref{Dicke_H}) we find,  to leading order, the quadratic Hamiltonian  
\begin{equation}\label{Dicke_H2}
H = \tilde{\omega}_0 \delta b^\dagger \delta b + \omega \delta a^\dagger \delta a + \tilde{\lambda} (\delta a + \delta a^\dagger) (\delta b + \delta b^\dagger) - \zeta (\delta b + \delta b^\dagger)^2,
\end{equation}
where 
\begin{IEEEeqnarray}{rCl}
    \tilde{\omega}_0 &=& \omega_0 - \lambda (\alpha + \bar{\alpha}) \frac{\tilde{\beta}_-}{\tilde{\beta}_+}, \\[0.2cm]
    \tilde{\lambda} &=& \lambda \tilde{\beta}_+ \bigg(1 - \frac{\tilde{\beta}_-^2}{\tilde{\beta}_+^2}\bigg),  \\[0.2cm]
    \zeta &=& \frac{\lambda(\alpha + \bar{\alpha})}{2} \frac{\tilde{\beta}_-}{\tilde{\beta}_+} \bigg( 1 + \frac{\tilde{\beta}_-^2}{2\tilde{\beta}_+^2}\bigg).
\end{IEEEeqnarray}

\subsection{Stabilization of the solution}

The Gaussianization procedure above explicitly already takes the limit $N\to \infty$. 
Because of this, it turns out that on order to obtain a stable steady-state, it is also necessary to add a small dissipation to $\delta b$. Here we do so in the simplest way possible, as a zero temperature dissipator. 
We therefore consider the evolution of the Gaussianized master equation
\begin{equation}
    \frac{d \rho}{d t} = -i [H,\rho] + 2\kappa \mathcal{D}[\delta a] + 2 \gamma \mathcal{D}[\delta b],
\end{equation}
where $D[L] = L\rho L^\dagger - \frac{1}{2} \{L^\dagger L, \rho\}$.
The value of $\gamma$ was actually  determined experimentally in~\cite{Brunelli2018ExperimentalSystems} and is more than six orders of magnitude smaller than $\kappa$.
One must therefore use a non-zero value, but the value itself can be arbitrarily small. 
In Fig.~\ref{fig:dicke} of the main text, we have used $\gamma = 10^{-3} \kappa$ simply to ensure numerical stability. 

\subsection{Lyapunov equation}

Once Gaussianized, we can  study the steady-state by solving for the second moments of $\delta a$ and $\delta b$.
Define  quadrature operators
\begin{equation}
    \delta q_b = \frac{\delta b + \delta b^\dagger}{\sqrt{2}} \qquad\quad \delta p_b = \frac{i}{\sqrt{2}} (\delta b^\dagger - \delta b),
\end{equation}
with identical definitions for $\delta q_a$ and $\delta p_a$. 
The Hamiltonian~(\ref{Dicke_H2}) then transforms to 
\begin{equation}\label{Dicke_H2_quadratures}
    H_2 = \frac{\tilde{\omega}_0}{2} (\delta q_b^2 + \delta p_b^2) + \frac{\omega}{2} (\delta q_a^2 + \delta p_a^2) + 2 \tilde{\lambda} \delta q_a \delta q_b - 2 \zeta \delta q_b^2. 
\end{equation}

Next define the Covariance Matrix (CM)  
\begin{equation}\label{SM_Dicke_CM}
    \sigma_{ij} = \frac{1}{2} \langle \{ R_i, R_j\} \rangle \qquad \bm{R} = (\delta q_b,\delta p_b, \delta q_a, \delta p_a). 
\end{equation}
Since both the Hamiltonian and the dissipator are Gaussian preserving, the dynamics of $\sigma$ is closed and described by a Lyapunov equation
\begin{equation}
    \frac{d \sigma}{d t} = A \sigma + \sigma A\trans + \mathcal{D},
\end{equation}
where
\begin{equation}
    A = 
    \begin{pmatrix}
    -\gamma                     &   \tilde{\omega}_0        & 0                     & 0 \\[0.2cm]
    4 \zeta - \tilde{\omega}_0  &   -\gamma                 & -2 \tilde{\lambda}    & 0 \\[0.2cm]
    0                           &   0                       &  -\kappa              & \omega\\[0.2cm]
    - 2 \tilde{\lambda}         &   0                       &   -\omega             &  -\kappa
    \end{pmatrix}
\end{equation}
and $\mathcal{D} = \text{diag}(\gamma,\gamma,\kappa,\kappa)$.

The assumption that the state of the system can be Gaussianized allows us to write down the Husimi function of the NESS, which has the form
\begin{equation}
    Q = \frac{1}{\pi \sqrt{|\sigma + \mathbb{I}_4/2|}} \exp\bigg\{ - \frac{1}{2} \bm{r}\trans (\sigma+ \mathbb{I}_4/2)^{-1} \bm{r}\bigg\}.
\end{equation}
where $\bm{r} = (x_b,y_b,x_a,y_a)$ are the phase space variables corresponding to the quadrature operators $\bm{R}$ in Eq.~(\ref{SM_Dicke_CM}) and $\mathbb{I}_4$ is the identity matrix of dimension 4.  
All integrals appearing in Eq.~(\ref{Pi_sep}) will then be Gaussian and can thus be  trivially computed. 

\end{document}